\documentclass[proceedings]{JHEP3} % 10pt is ignored!

\JHEPspecialurl{http://jhep.sissa.it/JOURNAL/JHEP3.tar.gz}

\usepackage{epsfig,multicol}

\newcommand\fverb{\setbox\pippobox=\hbox\bgroup\verb}
\newcommand\fverbdo{\egroup\medskip\noindent%
            \fbox{\unhbox\pippobox}\ }
\newcommand\fverbit{\egroup\item[\fbox{\unhbox\pippobox}]}
\newbox\pippobox
\def\a{\alpha}
\def\b{\beta}

\def\l{\lambda}

\def\s{\sigma}

\def\be{\begin{equation}}
\def\ee{\end{equation}}
\def\beq{\begin{eqnarray}}
\def\eeq{\end{eqnarray}}

\title{Cosmological billiards and oxidation}

\author{S. de Buyl$^a$\thanks{Aspirant du Fonds National de la
Recherche Scientifique, Belgique}~,
        M. Henneaux$^{a,b}$, B. Julia$^c$ and Louis Paulot$^a$\\
    $^a$ Physique Th\'eorique et Math\'ematique, Universit\'e Libre de Bruxelles, \\
    \hspace{.15cm} C.P. 231; B-1050, Bruxelles, Belgium\\
    $^b$ Centro de Estudios Cient\'{\i}ficos,\\ \hspace{.15cm}  Casilla 1469, Valdivia, Chile \\
$^c$ Laboratoire de Physique Th\'eorique de l'Ecole Normale
Sup\'erieure, \\ \hspace{.15cm}  24, rue Lhomond, F-75231 Paris
CEDEX 05,
France\\
    E-mail: \email{sdebuyl@ulb.ac.be}, \email{henneaux@ulb.ac.be},
    \email{Bernard.Julia@lpt.ens.fr},
    \email{Louis.Paulot@ulb.ac.be}}

\preprint{\hepth{0312251}}  % OR: \preprint{Aaaa/Mm/Yy\\Aaa-aa/Nnnnnn}
\abstract{We show how the properties of the cosmological billiards
provide useful information (spacetime dimension and $p$-form
spectrum) on the oxidation endpoint of the oxidation sequence of
gravitational theories.  We compare this approach to the  other
available methods: $GL(n,R)$ subgroups and the superalgebras of
dualities.}

\begin{document}

%\maketitle  IS IGNORED %%%%%%%%%%%

\section{Introduction}
It has been shown recently that the dynamics of gravitational
theories can be described, in the vicinity of a spacelike
singularity, as a billiard motion in a region of hyperbolic space
bounded by hyperplanes \cite{BKL}-\cite{DHN}. For gravitational
theories that reduce upon toroidal compactification to $D= 3$
dimensions to $3$-dimensional gravity coupled to a symmetric space
non-linear $\sigma$-model, the billiard turns out to have
remarkable regularity properties: it is the fundamental Weyl
chamber of the overextension \cite{J,GO} of the symmetry group
${\cal U}_3$ \cite{DH2}-\cite{DdBHS} that emerges in $3$
dimensions or, when the real form ${\cal U}_3$ is not split, of
its maximal split subalgebra \cite{HJ}.

An interesting feature of the billiard region is that it is
invariant under toroidal dimensional reduction to any dimension
$D\geq 3$ \cite{DdBHS}. Knowing the billiard region in $D=3$
dimensions can then be used as a tool to determine the possible
higher dimensional parents of the theory.  This problem of going
up in dimension is known as the oxidation or disintegration
problem and goes back to the early days of supergravity and has
been studied repeatedly \cite{J,J', J''}. It has been thoroughly
investigated recently by means of group theory techniques
\cite{Keur1}-\cite{Keur3}. We show here that the billiard approach
gives direct information on possible obstructions to oxidation
and, when oxidation is possible, restricts efficiently  the
maximal dimension(s) of oxidation as well as determines the full
$p$-form spectrum of the maximally oxidized theory.  We consider
both split and non-split ${\cal U}_3$-groups.  For non maximally
non-compact groups (ie non-split), this approach is complementary
to the general method of general linear subgroups of \cite{J''}
known as the ``A-chain'' method. We also point out that the same
information can be extracted from the superalgebra approach to the
problem \cite{HLJP1}-\cite{HLJP3}.

\section{Walls associated with simple roots and oxidation}

The walls bounding the billiards have different origins
\cite{DH2,DHN}: they can be symmetry or curvature walls, related
to the Einstein-Hilbert action; or they can be $p$-form walls
(electric or magnetic), related to the $p$-form part of the
action.  The key to the derivation of the oxidation constraints is
to investigate how the billiard walls behave upon dimensional
reduction. Although the billiard region is invariant, the formal
origin of the walls may  change. E.g., a symmetry wall in $D$
dimensions may appear as a $1$-form wall associated with the
Kaluza-Klein graviphoton(s) in lower dimensions \cite{DH2}.

The translation rules relating walls in higher dimensions to walls
in lower dimensions have been worked out in \cite{DdBHS,HJ}.  To
recall them, we denote by  $\bar{B}$ the restricted root system of
the real form ${\cal U}_3$.  This restricted root system may be
reduced or non reduced, in the latter case it is of $BC_r$-type
\cite{Helgason}. The billiard region is the fundamental Weyl
chamber of the overextension $\bar{B}^{\wedge \wedge}$ \cite{HJ}
obtained by adding to the simple roots of the finite root system
$\bar{B}$, the affine root $\a_0$ and the ``overextended root"
$\a_{-1}$, which is attached to the affine root with a single
link. There are at most two different root lengths, except when
the underlying finite root system is non reduced, in which case
one has three different root lengths.  The highest root is always
a long root (``very long" root in the $BC_r$-case) \cite{HJ}.

{}From the point of view of the three-dimensional action
(Einstein-Hilbert action + ${\cal U}_3$-non linear $\sigma$-model
action), these roots have the following interpretation\footnote{As
in our previous work, we denote the logarithmic scale factors of
the spatial metric by $\b^1$ and $\b^2$.  The scalar fields of the
$\s$-model Iwasawa-split into the dilatons $\phi^\Delta$ ($\Delta
= 1, \cdots, r$ where $r$ is the real rank) and the axions
($0$-forms) $\chi^A$.}: (i) the overextended root is the symmetry
wall $\b^2 - \b^1$ and involves only gravitational variables; (ii)
the affine root corresponds to the dominant magnetic wall $\b^1 -
\theta(\phi)$ where $\theta(\phi)$ is the highest root of
${\bar{B}}$; it involves both the dilatons and the gravitational
variable $\b^1$; and (iii) the simple roots of $\bar{B}$
correspond to the dominant electric walls defined by the $0$-forms
(axions) and involve the dilatons only.

In $D= d+1  >3 $ dimensions, these walls appear as follows. (i)
The overextended root remains a symmetry wall\footnote{Scale
factors get redefined as one changes the spacetime dimension but
we shall keep the same notation since the context is in each case
clear.}, namely $\b^{d} - \b^{d-1}$; (ii) The affine root becomes
the symmetry wall $\b^{d-1} - \b^{d-2}$. (iii) The (long) roots of
${\bar B}$ connected to the affine root by a chain of single links
form a chain of symmetry walls $\b^{d-2} - \b^{d-3}$, $\cdots$,
$\b^2 - \b^1$.  The other roots correspond to $p$-form walls of
the form $\b^1 + \cdots +  \b^p + \frac{1}{2} \sum_\a \l^{(p)}_\a
\phi^\a $ (electric walls) or $\b^1 + \cdots +  \b^{d-p-1} -
\frac{1}{2} \sum_\a \l^{(p)}_\a \phi^\a$ (magnetic
walls)\footnote{For pure gravity, one curvature wall, namely $2
\b^1 + \b^2 + \cdots + \b^{d-2}$ defines also a simple root.  This
case is covered in \cite{DHJN}.}.  In these equations, the
$\l^{(p)}_\a$ are the couplings to the dilatons $\phi^\a$ present
in $D$ dimensions (if any).

\section{Obstructions to oxidation}
A crucial feature of the gravitational walls (symmetry and
curvature walls) is that they are non degenerate.  By contrast,
the $p$-form walls may come with a non trivial multiplicity since
one can include many $p$-forms with the same degree and same
dilaton couplings.

It follows from this observation that a ``necessary condition''
for oxidability of a $3$-dimensional theory to $D>3$ dimensions is
that the (very) long roots of $\bar{B}$ must be non degenerate,
since the affine root, which is the magnetic wall $\b^1 -
\theta(\phi^\Delta)$ in three dimensions, becomes a symmetry wall
in $D>3$ dimensions (and the degeneracy of $\b^1 -
\theta(\phi^\Delta)$ is the degeneracy of the (very) long highest
root $\theta(\phi^\Delta)$).  This necessary condition on the
highest root was spelled out in \cite{HJ}. It provides an
obstruction to oxidation to $D>3$ dimensions for the following
cases (see \cite{Helgason}) for conventions):

\vspace{.2cm} \noindent  $A\; II$ ($SU^*(2n)/Sp(n)$) for which $
m_\theta = 4$; $B\; II$ and $D\; II$ ($SO(p,1)$) for which
$m_\theta = p -1$; $C\; II$ ($Sp(p,q)/Sp(p)\times Sp(q))$ for
which $m_\theta = 3$; $E\; IV$ ($E_{6 (-28)}/F_4$) for which
$m_\theta = 8$; and $F\; II$ ($F_{4(-20)}/SO(9)$) for which
$m_\theta = 7$. Here, $m_\theta$ is the multiplicity of the
highest root $\theta$.  That these non-split theories cannot be
oxidized agrees with the findings of \cite{BGM,Keur2}.

It might help to notice that this list is precisely the list of
real forms whose Satake diagrams have a ``compact'' Cartan
generator at the simple root(s) that connects (or connect) to the
affinizing root or equivalently that is not orthogonal to the most
negative root of the simple Lie algebra under consideration.
Affinization occurs in 2 dimensions; in the build-up of the
$SL(D_{max}-3,R) \times R $ chain one would expect the dilation of
the fourth coordinate to be a noncompact Cartan generator at that
vertex as soon as $D_{max} \ge 4$  \cite{J''}.

Conversely the fact that the root before the affinizing one is
noncompact, to use approximate language, seems to imply that the
affine root is noncompact too and can participate in a
$GL(D_{max}-2,R)$ group of duality symmetries \cite{J''}.

Finally, we note that the $C\; II$ (with $q=2$) and $F\; II$ cases
have a supersymmetric extension with an odd number of
$3$-dimensional supersymmetries ($N= 5$ and $N=9$, respectively)
\cite{dWN}. These supersymmetries cannot come from four dimensions
since a four-dimensional theory yields an even number of
supersymmetries in $D=3$. Our obstruction to oxidation given above
does not rely on supersymmetry, however,  and prevents even
non-supersymmetric ``parents".

\section{Maximal dimensions}
When the theory can be oxidized, one can infer the maximal
dimension(s) in which it can be formulated from the rules recalled
above: this is $D_{max} = k + 2$ where $k$ is the length of the
symmetry wall chain, i.e. the length of the single-linked chain of
nodes in the Dynkin diagram of the restricted root system
$\bar{B}^{\wedge \wedge}$ that can be constructed starting from
the overextended root, without loop (hereafter called ``A-chain").
If there is a fork, each branch yields an independent maximal
dimension. By mere inspection of the tables and Satake diagrams in
\cite{Helgason} and of the Dynkin diagrams of the overextensions
given e.g. in \cite{DdBHS,HJ}, one easily gets the results
collected in the following table:

\begin{center}
\begin{tabular}{|l|l|c|}
\hline  Class & Noncompact Symmetric Space D=3 & $D_{max}$ \\
\hline  $A \; I$ &$SL(r+1)/SO(r+1)$ & $r+3$  \\

\hline  $A\; III$ and
$A\; IV$ &$SU(p,q)/(S(U_p \times U_q))$  & $4$  \\
\hline  $B\; I$ and $D\; I$ &$SO(p,q) / (SO(p) \times SO(q))$,
$p \geq q >1$  & $q+2$ and $6$ ($q \geq 3$) \\
\hline $C\; I$ &$Sp(n,R)/U(n)$  & $4$ \\
\hline $D\; III$& $SO^*(2n)/U(n)$  & $4$  \\
\hline $E\; I$ & $E_{6(6)}/Sp(4)$ &  $8$ \\
 \hline $E\; II$&
$E_{6(2)}/(SU(6) \times SU(2))$& $6$ \\
\hline $E\; III$&
$E_{6(-14)}/(SO(10) \times U(1))$& 4\\
\hline $E\; V$&
$E_{7(7)}/SU(8)$ & $8$ and $10$ \\
\hline $E\;
VI$ &$E_{7 (-5)}/(SO(12) \times SU(2))$ & 6 \\
\hline $E\; VII$ & $E_{7 (-25)}/ (E_{6 (-78)} \times U(1))$ & 4 \\
\hline $E\; VIII$ & $E_{8(8)}/SO(16) $ & 10 and 11 \\
\hline $E\; IX$ & $E_{8(-24)}/(E_{7(-133)} \times SU(2))$ & 6 \\
\hline  $F\; I$ &$F_{4(4)}/(Sp(3) \times SU(2))$ &6\\
\hline   $G$ & $G_{2(2)}/SO(4)$& 5 \\

\hline
\end{tabular}
\end{center}

In all maximal oxidation dimensions, a theory which correctly
reduces to $D=3$ actually exists.  These theories are listed in
\cite{Keur2} in terms of previously constructed models
(\cite{BGM,CJLP,others}).

It is interesting to observe that the same conclusions can be
reached from other approaches to the hidden symmetries of
gravitational theories.  In the approach where the forms are
associated with the roots of a superalgebra
\cite{HLJP1}-\cite{HLJP3}, one needs only replace the couple of
the affine and extended roots by the $D=3$ fermionic root, which
must be of multiplicity one if the theory comes from a higher
($D>3$) dimension; the rest of the discussion proceeds in the same
way. A real chain $\mathfrak{sl}(n|1)$ in the $D=3$ superalgebra
(with long roots) allows to reach dimension $D=3+n$. Reduction on
an $n$-torus gives indeed $n$ odd roots $E_i$ ($i=1,...,n$) which
combine into a $\mathfrak{sl}(n|1)$ algebra, with simple roots
$E_0-E_1, \ldots, E_{n-1}-E_n$ (even) and $E_n$ (odd). When
considering the restricted roots system, as we do in the billiard
approach, it corresponds to a nondegenerate $\mathfrak{sl}(n|1)$
chain.

In the approach where it is the triple extension that appears
\cite{EHTW1,EHW2}, one gets the same conclusion because the
overextended part is identical.

\section{$p$-form spectrum and oxidation}
The billiard provides also information on the $p$-forms that must
be present in the maximal oxidation dimension because one knows
how the $p$-form walls must connect to the A-chain.  A simple
electric $p$-form wall connects to the $p$-th root in the A-chain
(counting now from the root at the end of the A-chain opposite to
the overextended root). A simple magnetic $p$-form wall connects
to the $(d-p-1)$-th root. One also knows the multiplicities.
Together with the Weyl group, this can be used to construct the
$p$-form spectrum and determine the dilaton couplings.

Rather than deriving explicitly all cases, we shall focus on $E\,
\; IX$, i.e., $E_{8(-24)}/(E_{7(-133)} \times SU(2))$, which has a
restricted root system of $F_4$-type

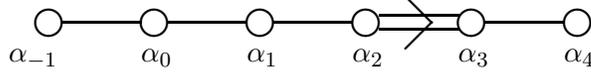
\begin{center}
\scalebox{1}{

\begin{picture}(220,20)

\thicklines \multiput(10,10)(40,0){6}{\circle{10}}
\multiput(15,10)(40,0){3}{\line(1,0){30}}
\put(175,10){\line(1,0){30}} \put(135,12.5){\line(1,0){30}}
\put(135,7.5){\line(1,0){30}} \put(145,0){\line(1,1){10}}

\put(145,20){\line(1,-1){10}} \put(-5,-5){$\alpha_{-1}$}
\put(45,-5){$\alpha_{0}$} \put(85,-5){$\alpha_{1}$}
\put(125,-5){$\alpha_{2}$}
\put(165,-5){$\alpha_{3}$}\put(205,-5){$\alpha_{4}$}
\end{picture}
}
\end{center}

The other cases are treated similarly and reproduce known results.
The root $\a_{-1}$ in the above Dynkin diagram is the overextended
root, the root $\a_0$ is the affine root, while $\a_1$ and $\a_2$
are the long roots of $F_4$. The roots $\a_{-1}, \a_0, \a_1, \a_2$
have multiplicity $1$. The roots $\a_3$ and $\a_4$ are the short
roots and have multiplicity $8$ \cite{Helgason}. The A-chain is
given by the roots $\a_{-1}, \a_0, \a_1, \a_2$, which read, in
$D_{max}=6$ dimensions, $\a_2 = \b^2 - \b^1$, $\a_1= \b^3 - \b^2$,
$\a_0 = \b^4 - \b^3$ and $\a_{-1} = \b^5 - \b^4$. Since $F^{\wedge
\wedge}_4$ has rank $6$ and since there is only five logarithmic
scale factors, one needs one dilaton.

Because the short root $\a_3$ is attached to the first root $\a_2$
of the A-chain, it corresponds to the electric wall of a $1$-form.
Requiring that the root has length squared equal to one (one has
$(\a_2 \vert \a_2) = 2$ and $(\a_2 \vert \a_3) = -1$) fixes the
dilaton coupling of the one-form to $\lambda^{(1+)} = - 1$, i.e.,
$\a_3 = \b^1 - (\phi/2)$. The last simple root $\a_4$ is not
attached to the A-chain: it corresponds therefore to an axion,
with dilaton coupling equal to $\lambda^{(0)} =  2$, i.e.,
electric $0$-form wall $\a_4 = \phi$. The degeneracy of the short
roots is $8$; hence, at this stage, we need eight $1$-forms and
eight $0$-forms.

This is not the entire spectrum of $p$-forms because we have only
accounted so far for the simple roots.  There are other positive
roots in the theory which correspond also to walls of the $p$-form
type, as can be seen by acting with the finite Weyl group of $F_4$
(= restricted Weyl group of ${\cal U}_3$) on the roots $\a_1$,
$\a_2$, $\a_3$ and $\a_4$. These roots must be included by
incorporating the corresponding $p$-forms.

By Weyl-reflecting the short root $\a_3 = \b^1 - (\phi/2)$ in
$\a_4$, we get the root $\b^1 + (\phi/2)$. This is also an
electric $1$-form wall, with degeneracy $8$. We thus need eight
further $1$-forms, with dilaton couplings $\lambda^{(1-)} =  1$.
We then observe that the symmetry wall reflection in $\b^2 - \b^1$
replaces $\b^1 - (\phi/2)$ by $\b^2 - (\phi/2)$. Reflecting this
root in $\b^1 + (\phi/2)$ yields $\b^1 + \b^2$, which is the
electric (= magnetic) wall of a chiral $2$-form with zero dilaton
coupling. This root is short, so again degenerate $8$ times: we
need eight such chiral $2$-forms (or four non-chiral $2$-forms
with zero dilaton couplings). Finally, by reflecting in $\a_3$ the
long root $\b^2 - \b^1$, we generate the long root $\b^1 + \b^2 -
\phi$. This is an electric wall for a $2$-form with dilaton
coupling $\lambda^{(2)} = - 2$, which we must include.  A
reflection in $\a_4$ yields $\b^1 + \b^2 + \phi$ but this is just
a magnetic root for the same $2$-form, so we do not need any
further additional field to get these walls in the Lagrangian.
Similarly, the short roots $\b^1 + \b^2 + \b^3 \pm (\phi/2) $
obtained by reflecting $\b^3 \pm (\phi/2)$ in $\b^1 + \b^2$ are
the magnetic walls of the $1$-forms and do not need new fields
either. [Note in passing the obvious misprints in formulas (6.21)
and (6.22) for the magnetic walls of the $2$-forms in
\cite{DdBHS}. (Also the magnetic walls of $G_2$ are miswritten
there).]

To summarize: by acting with the finite Weyl group of $F_4$ on the
simple roots $\a_1$, $\a_2$, $\a_3$ and $\a_4$, we get all the
roots of $F_4$.  The positive roots, which must have a term in the
Lagrangian, are $\b^i - \b^j$ ($i>j)$ (long), $\phi$ (short),
$\b^i \pm (\phi/2)$ (short), $\b^i + \b^j$ ($i <j$) (short), $\b^i
+ \b^j \pm \phi$ ($i <j$) (long), $\b^i + \b^j + \b^k \pm
(\phi/2)$ ($i<j<k$) (short) and $2 \b^i + \b^j + \b^k$ ($i \not=
j$, $i \not=k$, $j<k$) (long). Here, $i,j,k \in \{1,2,3\}$.  These
are all the $24$ positive roots of $F_4$.  They are all accounted
for by the $p$-form walls, except the last ones, which are
curvature walls following from the Einstein-Hilbert action (and
the symmetry walls).  By acting with the $F_4^{\wedge
\wedge}$-Weyl reflections associated with the other symmetry
walls, one covariantizes the above expressions (i.e., one
generates the same walls but with $i,j,k \in \{1,2,3,4,5\}$).
Continuing, i.e., acting with the other elements of the infinite
Weyl group of $F_4^{\wedge \wedge}$, one generates new walls but
these are not of the $p$-form type because they necessarily
contain a $\b^i$ with a coefficient at least 2 (except the
magnetic walls $\b^i + \b^j + \b^k + \b^m - \phi$ ($i<j<k<m$) of
the $0$-form and the null magnetic walls $\b^i + \b^j + \b^k +
\b^m$ of the dilaton, but these are already accounted for). All
the $p$-form roots have been exhausted.

The superalgebra approach gives an alternative procedure to
generate the $p$-spectrum from the Borcherds-Chevalley-Serre
relations and also provides detailed information on the
Chern-Simons terms in the Lagrangian (as the group theory approach
\cite{Keur1}-\cite{Keur3} and the principles given in \cite{EHW2}
do). See \cite{HLJP1,HLJP3} for the split case and
\cite{HLJP2,HLJP3} for the non split case. In this formalism,
there are roots associated to scalars but also to nonreduced
$p$-forms. If $\alpha$ is a root associated to a $p$-form in the
superalgebra of the $D$-dimensional theory, one gets in addition
in lower dimensions roots like $\alpha-E_i$ (corresponding to a
$(p-1)$-form), $\alpha-E_i-E_j$ ($(p-2)$-form)\ldots{} All
original roots remain in the spectrum, and the oxidation procedure
computes precisely those roots which are present in the oxidised
theory.

If we consider again the \emph{E IX} example, the superalgebra of
the $D=3$ theory is given by the following Satake diagram:
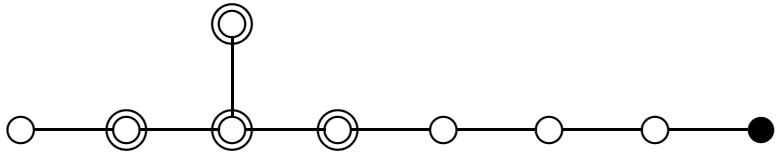
\begin{center}
\scalebox{1}{
\begin{picture}(300,60)
\thicklines \multiput(10,10)(40,0){7}{\circle{10}}
\put(290,10){\circle*{10}} \put(90,50){\circle{10}}
\multiput(15,10)(40,0){7}{\line(1,0){30}}
\put(90,15){\line(0,1){30}} \multiput(50,10)(40,0){3}{\circle{15}}
\put(90,50){\circle{15}}
\end{picture}
}
\end{center}
where white dots are even simple roots of degree 0 and the black
one an odd $\mathfrak{sl}(1|1)$ one of degree 1; the circled roots
are those which are antiinvariant under the Cartan involution
defining the real form. One sees on this diagram an
$\mathfrak{sl}(3|1)$ chain that is invariant under the Cartan
involution (its 3 simple roots are the rightmost ones which are
orthogonal to, ie disconnected from, the antiinvariant ones). It
means that we can oxidise the model up to $D=3+3=6$.

The superalgebra of the maximal oxidised theory can be easily
deduced; it is given by the following diagram:
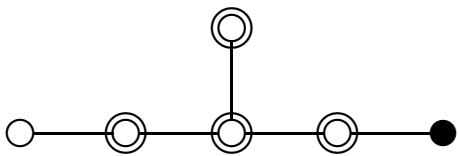
\begin{center}
\scalebox{1}{
\begin{picture}(180,60)
\thicklines \multiput(10,10)(40,0){4}{\circle{10}}
\put(170,10){\circle*{10}} \put(90,50){\circle{10}}
\multiput(15,10)(40,0){4}{\line(1,0){30}}
\put(90,15){\line(0,1){30}} \multiput(50,10)(40,0){3}{\circle{15}}
\put(90,50){\circle{15}}
\end{picture}
}
\end{center}
We see that the real rank of this superalgebra is 2, from what we
know there is $2-1=1$ dilaton in the spectrum. We can also compute
that there are 10 degree zero positive roots, so 8 0-forms; there
are also 16 degree 1 roots, obtained by acting on the odd simple
root with the Weyl group of $D_5$, which gives us 16 1-forms.
There are 10 degree 2 roots, but related by duality: there are
only 5 dynamical 2-forms (2 chiral 2-forms count as a single
ordinary 2-form). In addition we find all magnetic duals of the
dilaton, 0- and 1-forms: 16 3-forms and 8+1 4-forms.

\section{Results and comments}
We have shown in this paper how the billiard analysis of
gravitational theories, related to Satake diagrams and restricted
root systems \cite{HJ},  provides useful information on their
oxidation endpoint both in the split and in the non-split cases.
Note, in particular, that our results suggest a 6-dimensional
formulation of magic square supergravity theories
\cite{Guna1,Guna2} since these have a $F_4$ restricted root
system, this was alrealdy announced a few years ago \cite{J1999}.
The real form $E \; IX$ with the $D=6$ spectrum displayed above
would in fact correspond to the octonion case.

It is noteworthy that the split real forms allow oxidation by at
least one dimension, to $D \geq 4$. One is stuck to $D=3$ only for
certain non-split real forms, because of the non trivial
multiplicities of the (very) long roots (which are always
non-degenerate in the split case). The constraint on
multiplicities provides a rather powerful insight. {}Finally, we
mention the recent paper \cite{Fre} which
also deals with ${\cal U}_3$, billiards and oxidation.

\acknowledgments

Marc Henneaux is grateful to the organizers of the 27th Johns
Hopkins workshop and of the 36th International Symposium
Ahrenshoop where he presented the results of the papers
\cite{DH1,DH2,DHN,DdBHS,HJ}.  The work of S.d.B., M.H. and L.P. is
supported in part by the ``Actions de Recherche Concert{\'e}es" of
the ``Direction de la Recherche Scientifique - Communaut{\'e}
Fran{\c c}aise de Belgique", by a ``P\^ole d'Attraction
Interuniversitaire" (Belgium) and by IISN-Belgium (convention
4.4505.86). Support from  the European Commission RTN programme
HPRN-CT-00131, in which S.d.B., M.H. and L.P. are associated to K.
U. Leuven, is also acknowledged.

\end{document}